\documentclass[12pt]{article}
\pdfoutput=1

 \usepackage{amsmath,amsfonts,amssymb,mathrsfs,graphicx,xcolor,ifthen,times}
\usepackage{hyperref}
\usepackage[normalem]{ulem}

\newenvironment{sciabstract}{%
\begin{quote} \bf}
{\end{quote}}
\newcounter{lastnote}

\graphicspath{{./}}

\newcommand{\be}{\begin{equation}}
\newcommand{\ee}{\end{equation}}
\newcommand{\beq}{\begin{eqnarray}}
\newcommand{\eeq}{\end{eqnarray}}

\def\t13{\mathrel{{\theta_{13}}}}
\def\y12{\mathrel{{\tan^2 \theta_{12}}}}
\def\c2{\mathrel{{\chi^2 }}}
\def\msun{\mathrel{{M_\odot }}}
\def\rsun{\mathrel{{R_\odot }}}

\def \sun	 	{$_{\odot}$}

\newcommand{\n}{neutrino}
\newcommand{\ns}{neutrinos}
\newcommand{\ic}{IceCube}
\newcommand{\td}{TDE}
\newcommand{\bh}{SMBH}
\newcommand{\tds}{TDEs}
\newcommand{\dsg}{AT2019dsg}
\newcommand{\icev}{IceCube-191001A}
\newcommand{\sft}{Swift J1644+57}

\newcommand{\mm}{multi-messenger}
\newcommand{\scw}{Schwarzschild}

\newcommand{\bi}{\begin{itemize}}
\newcommand{\ei}{\end{itemize}}

\newcommand{\eg}{{\it e.g.}}

\newcommand{\eq}{Eq.}

\newcommand{\equ}[1]{\eq~(\ref{equ:#1})}

\newcommand{\apj}{Astrophys. J.}
\newcommand{\apjs}{Astrophys. J. Suppl.}

\author
{Walter Winter$^1$, Cecilia Lunardini$^2$ 
}

\date{\today}

\begin{document}

\title{A concordance scenario for the observed neutrino from a Tidal Disruption Event}


\baselineskip24pt

\maketitle 

\begin{sciabstract}
During a tidal disruption event, a star is torn apart by the tidal
forces of a supermassive black hole, with about 50\% of the star's mass
eventually accreted by the black hole. The resulting flare can, in
extreme cases of super-Eddington mass accretion, result in a
relativistic jet \cite{Hills75,Rees:1988bf,Lacy82,Phinney89}. While tidal disruption events have been
theoretically proposed as sources of high-energy cosmic rays \cite{Farrar:2008ex,Farrar:2014yla} and
neutrinos \cite{Wang:2011ip, Wang:2015mmh,Dai:2016gtz,Senno:2016bso,Lunardini:2016xwi,Guepin:2017abw,Biehl:2017hnb,Hayasaki:2019kjy}, stacking searches indicate that their contribution to
the diffuse extragalactic neutrino flux is very low \cite{Stein:2019ivm}. However, a
recent association of a track-like astrophysical neutrino
(IceCube-191001A; \cite{2019GCN.25913....1I}) with a tidal disruption event (AT2019dsg; \cite{Stein:2020tem})
indicates that some tidal disruption events can accelerate cosmic rays
to PeV energies. Here we introduce a phenomenological concordance
scenario with a relativistic jet to explain this association: an
expanding cocoon progressively obscures the X-rays emitted by the
accretion disk, while at the same time providing a sufficiently intense
external target of back-scattered X-rays for the production of neutrinos
via proton-photon interactions. We also reproduce the delay (relative to
the peak) of the neutrino emission by scaling the production radius with
the black body radius. Our energetics and assumptions for the jet and
the cocoon are compatible with expectations from numerical simulations
of tidal disruption events.  
\end{sciabstract}

\maketitle


On October 1st, 2019, a track-like astrophysical neutrino (named \icev ) was detected \cite{2019GCN.25913....1I};  a dedicated \mm\ follow up program revealed the  tidal disruption event \dsg\ as a  candidate source, with a $p$-value of 0.2\% to 0.5\% of random association~\cite{Stein:2020tem}, corresponding to ${\sim 3 \sigma}$ significance. The neutrino followed the peak of the \dsg\ lightcurve by ${t-t_{\mathrm{peak}}=154}$ days and had a most likely energy ${E\simeq 0.2\, \mathrm{PeV}}$ {[}Ref. \cite{2019GCN.25913....1I} and links therein{]}. Its observation reveals a new class of cosmic ray sources, as it indicates that some tidal disruption events can accelerate cosmic rays to PeV energies.

\begin{figure*}[t]
\begin{center}
    \includegraphics[width=0.7\textwidth]{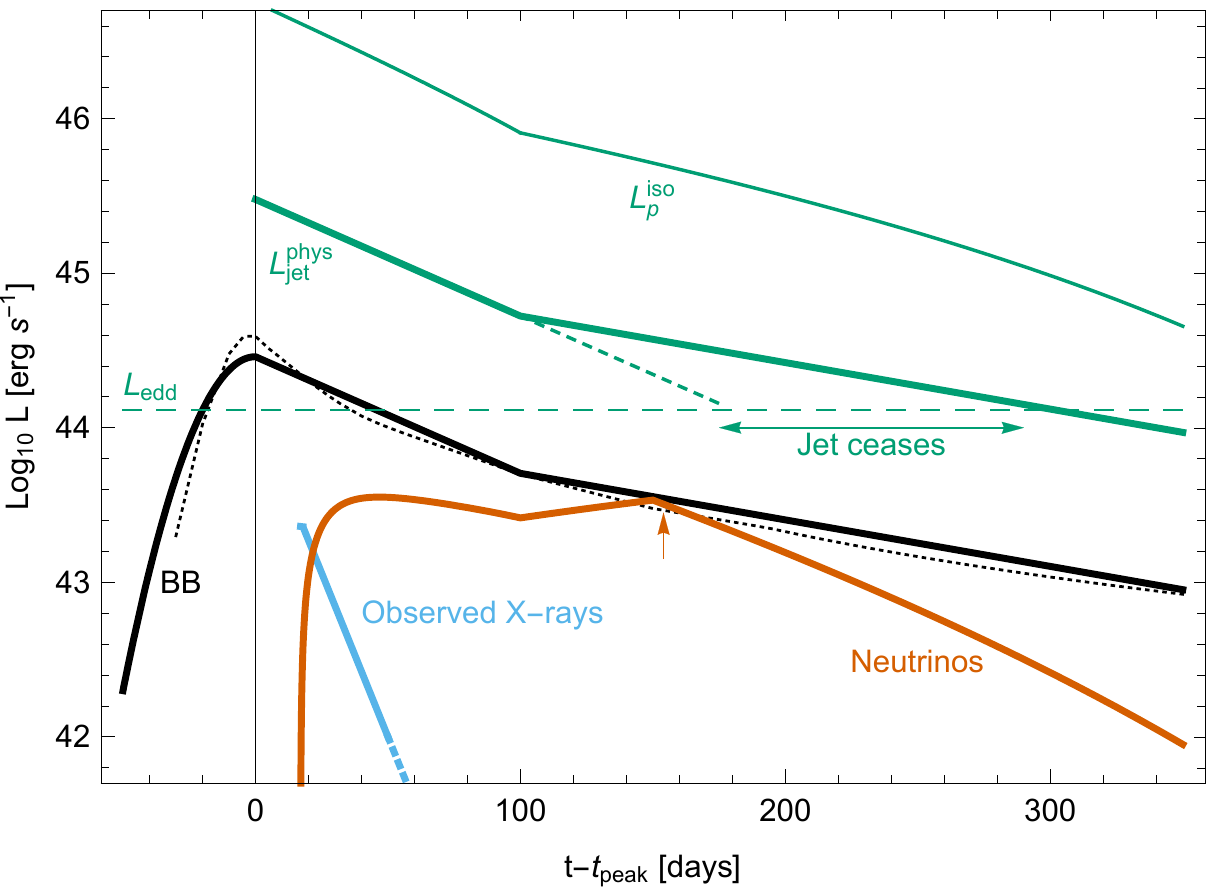}
\end{center}
\caption{{\bf Time evolution of different luminosities in the jetted TDE model.} The labels are directly given next to the curves, see main text for definitions. The neutrino luminosity is a result of our work, whereas the other curves are input quantities of the model. Thick black and blue (solid) curves 
(the latter starting at $t-t_{\mathrm{peak}}\simeq 17$ days, reflecting the lack of X-ray data prior to that point)
are chosen to roughly follow data \cite{vanVelzen:2020cwu}; the dotted black curve represents a power law fit from \cite{vanVelzen:2020cwu} which has a fixed late-time $t^{-5/3}$ behavior.   All shown luminosities are isotropic-equivalent, and refer to the source/engine frame, except for $L_{\mathrm{Edd}}$ and $L_{\mathrm{jet}}^{\mathrm{phys}}$. 
The vertical arrow marks the arrival time of the observed \n\ event. 
\label{fig:lumi_jetted}
}
\end{figure*}

The Tidal Disruption Event (TDE)  \dsg\ is located at redshift $z \simeq 0.05$, or luminosity distance $d_L \simeq 230$ Mpc. It was discovered in the Optical-UV bands by the Zwicky Transient Facility (ZTF) on  April 9th, 2019 \cite{vanVelzen:2020cwu}, and it reached its luminosity peak in this band on April 30th, 2019 ($t_{\mathrm{peak}}=58603$ MJD).  Several follow up observations were conducted in the optical-UV \cite{vanVelzen:2020cwu}, radio \cite{2019ATel12960....1P,2019ATel12798....1S,Stein:2020tem} and X-ray \cite{2019ATel12777....1P,Stein:2020tem,2019ATel12825....1P} bands, the latter starting at $t-t_{\mathrm{peak}}=17$ days.
The picture that emerged from the observations shows a several months-long flare, with black body spectra observed in both the optical-UV (temperature  $T_{\mathrm{BB}}=38900 \, \mathrm{K}$, photospheric radius $R_{\mathrm{BB}}\simeq 5\, 10^{14}\mathrm{cm}$) and X-ray ($T_X \sim 0.06$ keV,  $R_X \sim 3-7~ 10^{11}$ cm) bands, and luminosities exponentially decaying over an (initial)  time scale of 57.5 days and 10.3 days starting at $L_{\mathrm{BB}}=2.88 \cdot 10^{44} \, \mathrm{erg \, s^{-1}}$ and $L_X\sim 2.5 ~10^{43}~\mathrm{erg \, s^{-1}}$, respectively; see Fig.~\ref{fig:lumi_jetted} (thick black and blue curves). The quoted X-ray luminosity is for an energy window  $[0.3-8]~\mathrm{keV}$,  whereas an X-ray luminosity $L_X\sim 4~ 10^{44}~\mathrm{erg \, s^{-1}}$  was found in \cite{2019ATel12825....1P} in the energy window $[0.1-10]~\mathrm{keV}$.  Instead, the luminosity in radio emission was approximately constant over a nearly 90 days period, with increasing radius of emission $R_{\mathrm{radio}}={\mathcal O}(10^{16})~{\mathrm{cm}}$~{[}Ref. \cite{Stein:2020tem}{]}. The radio emission  has been interpreted as an indication for a mildly relativistic outflow present over the timescale of the neutrino event. Furthermore optical polarimetry observations of this TDE cannot be uniquely interpreted, and may provide some hint for a relativistic jet
\cite{Lee_2020}. A further noteworthy element is that out of the 17 \tds\ in the ZTF sample only four were found to have a counterpart in X-rays; of these, \dsg\ was the one with the highest \emph{sustained} (over  several days) X-ray luminosity. 

\begin{figure*}[tp]
\begin{center}
    \includegraphics[width=0.99\textwidth]{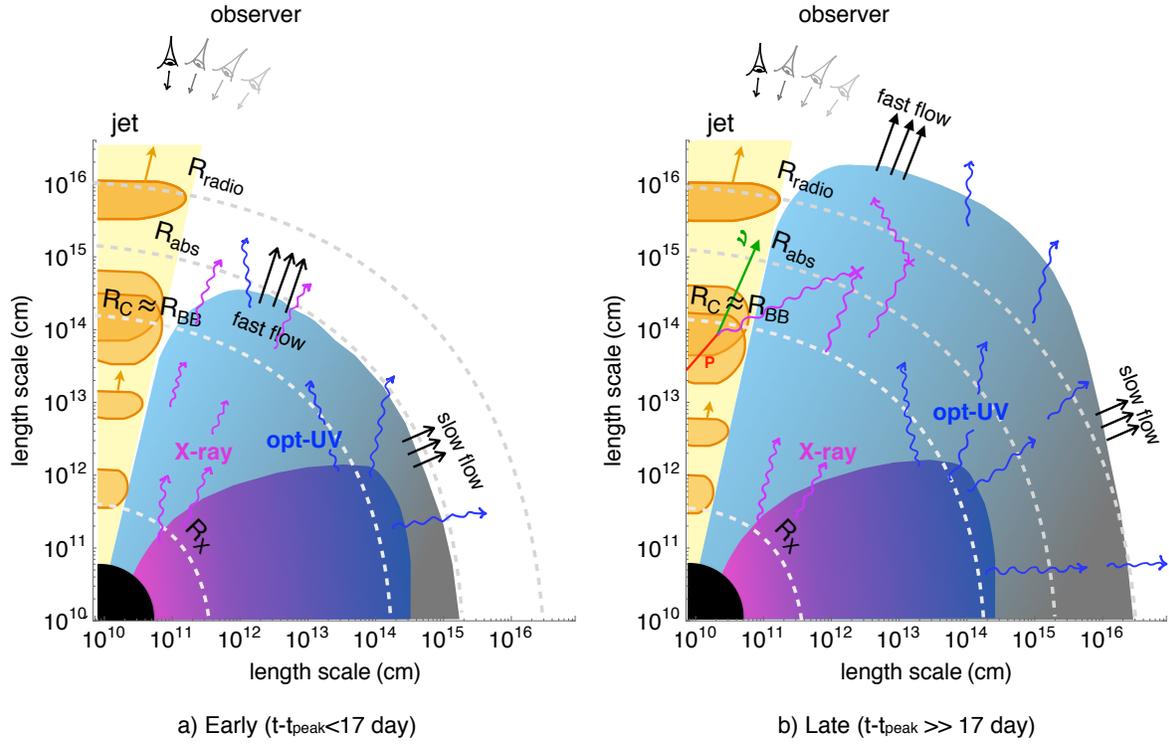} 
\end{center}
\caption{{\bf Illustration of the evolution of the TDE outflow in the concordance scenario.} Here two time periods are shown: a) early, where X-rays can efficiently escape, and b) late, where X-rays are absorbed/back-scattered into the jet frame. The relevant length scales  are marked as dashed lines to guide the eye (order magnitude only; as follows: $R_X$: X-rays photosphere radius, $R_{\mathrm{BB}}$: Black Body radius, $R_{\mathrm{radio}}$: radio emission region, $R_{\mathrm{abs}}$: X-ray mean free path, $R_C$: neutrino production radius). The anticipated direction of the observer is shown as well.  
 The proton acceleration and neutrino production is expected to happen at $R_C \sim R_{\mathrm{BB}}$, where plasma shells collide and shocks form. 
The flow expands slightly faster in the direction of the jet axis.
\label{fig:cartoon}
}
\end{figure*}

In this study we propose a coherent, ``concordance" framework of a (dark or hidden) jetted \td,  which is consistent with the unified model (based on magnetohydrodynamical simulations) of Dai, McKinney, Roth, Ramirez-Ruiz and Miller \cite{Dai:2018jbr}; see Methods for details. The framework describes the \n\ energy and arrival time -- where the latter is somewhat a challenge -- considering the overall decreasing trend of the \mm\ luminosities, see Fig.
~\ref{fig:lumi_jetted} (thick solid black and blue curves). A schematic concept is given in Fig.~\ref{fig:cartoon}. At early times, the X-rays are visible for the observer, whose line of sight is on (or close to) the jet axis. Plasma shells collide at radial distance $R_C$, where (internal) shocks form, leading to proton acceleration.
We assume that absorption by the expanding outflow causes  the exponential decay of the observed X-ray flux  (Fig.~\ref{fig:lumi_jetted}, thick blue curve).  Then, the same effect leads to photons isotropizing on the same timescale, which are back-scattered and Doppler-boosted into the jet frame where they serve as targets the neutrino production (right panel of Fig.~\ref{fig:cartoon}).  The physical jet power is taken from \cite{Dai:2018jbr}, and is assumed to follow the black body luminosity (Fig.~\ref{fig:lumi_jetted}, thick green curve). The jet ceases when the physical power drops below the Eddington luminosity. We assume that $R_C$ evolves similarly to  $R_{\mathrm{BB}}$, which is observed to decrease slightly over time.  A decreasing $R_C$ implies an increasingly compact collision region, and thus an increase of the neutrino production efficiency $\propto R_C^{-2}$ at late times. Similarly, a larger $R_C$, such as it may be expected for a larger \bh\ mass, would lead to a decreased neutrino production efficiency.

The result for the time evolution of the neutrino luminosity, $L_\nu$, is shown in Fig. \ref{fig:lumi_jetted} (red curve). 
Its initial rise and peak, at $t - t_{\mathrm{peak}}\sim$ 30-70 days, follows the isotropized target X-ray flux. 
At later times the evolution of $L_\nu$ is mostly 
determined by the competition of the decreasing proton injection and target photon densities (driving a decrease of the neutrino flux) and the decreasing production radius $R_C$ enhancing the neutrino flux.  
As a consequence of this interplay, the \n\ luminosity has a second peak
at $t - t_{\mathrm{peak}}\sim$ 130-170 days.
The late time luminosity revival could contribute to explaining the observed detection of one \n\ at $t - t_{\mathrm{peak}}=154$ days.  
Eventually, after the revival, $L_\nu$  undergoes a sharp drop  from the jet cessation or the $R_C$ stagnation (see Methods, \equ{rscale}).

\begin{figure}[tbp]
\begin{center}
    \includegraphics[width=0.6\textwidth]{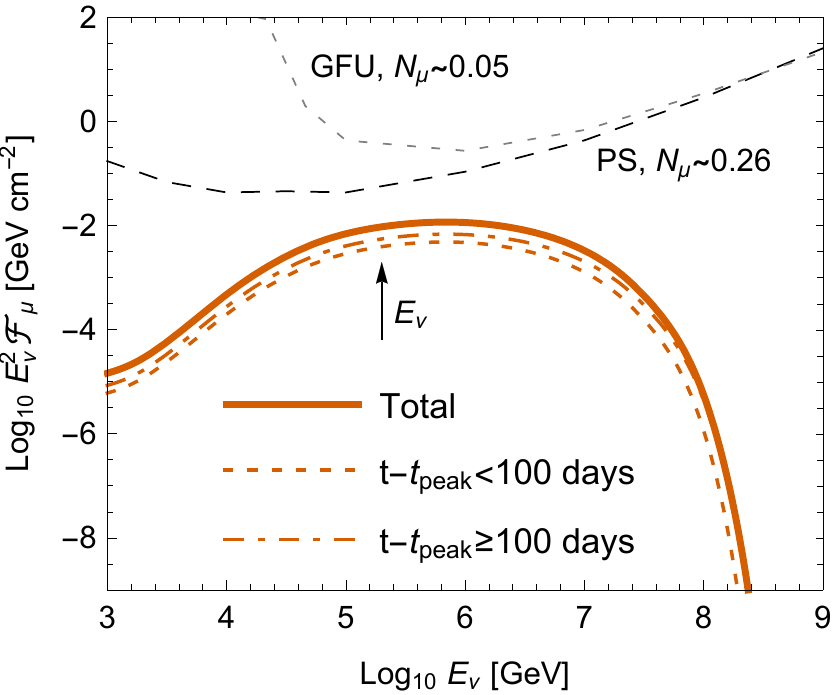}
\end{center}
\caption{{\bf Predicted neutrino fluence for the jetted TDE model.} The curves are computed as integrated muon neutrino and anti-neutrino fluence (including flavor mixing). The integrated contributions from early (red dashed) and late times (red dot-dashed)  are shown separately. In comparison, the differential limits and predicted event rates using the gamma-ray follow-up (GFU, \cite{Blaufuss:2019fgv}) and point source (PS, \cite{Aartsen:2014cva})  effective areas for the declination of AT2019dsg are shown; the likely neutrino energy is taken from \cite{Stein:2020tem}.
Here the differential limit is given by $E_\nu^2 \mathcal{F}_\mu^{\mathrm{DL}} =  E_\nu/(A_{\mathrm{eff}}(E_\nu) \, \mathrm{ ln } 10)$, which implies that following the differential limit curve precisely for one order of magnitude in energy yields one neutrino event. 
 }
\label{fig:fl_jetted}
\end{figure}

Fig.~\ref{fig:fl_jetted} shows the predicted neutrino fluence, $\mathcal{F}_\mu$, as well as two differential limits on the same quantity (see caption), for comparison. 
Compared to other cases of proton scattering on thermal X-rays, the \n\ energy spectrum is relatively wide here due to
multi-pion processes dominating the neutrino production (see, e.g., \cite{Hummer:2010vx} for a similar case). 
The most likely value of the \n\ energy ($E_\nu\sim 0.2$ PeV, with a large uncertainty allowing up to one order of magnitude larger values), falls near the maximum of the fluence. 

The total, time-integrated number of events predicted in IceCube depends on the effective area used. We find $N_\nu\simeq 0.26$ when using the point source effective area, (which applies to a transient point source analysis), and $N_\nu\simeq 0.05$ when using the gamma-ray follow-up effective area, which includes the probability that the alert system is triggered. 
Note that the observation of one event is well compatible with $N_\nu \ll 1$, due to Poissonian statistics, and due to the Eddington bias \cite{Strotjohann:2018ufz}.  From Fig.~\ref{fig:fl_jetted}, we also observe that the early- and late-term contributions to the total fluence are comparable, which implies that a neutrino detection $\sim$150 days after the peak is plausible. 

Let us discuss our proposed jet scenario  in the broader context. 
Compared to the best known jetted \td, \sft, \dsg\ is very different: it is $\sim 10^3$ times less powerful (from the observer's point of view) in X-rays, and its X-ray spectrum is thermal, in contrast with the non-thermal spectrum of \sft,
which was interpreted as signature of a jet. Therefore,  the existence of a jet in \dsg\ might be less obvious, and leaves some open questions.

One of these is how to reconcile the non-observation of (non-thermal) jet signatures in X-rays and gamma rays 
-- at least gamma rays in the 0.1--1 PeV energy range, which are a direct counterpart of the \ns\  --
with the expectation that the jet should be able to break out of a surrounding envelope material~\cite{Wang:2015mmh} and therefore should not be completely hidden.  

Currently, the sparseness of the data from \dsg\ (\eg, gamma-ray limits are relatively weak, see~\cite{Stein:2020tem}), does not allow for a clear description of the electromagnetic spectrum from the jet over a wide energy range.
With regard to the microphysics in the jet, there is no evidence for non-thermal signatures from accelerated electrons, such as synchrotron radiation and inverse Compton scattering. This fact may indicate a relatively high baryonic loading (energy in protons versus electromagnetic radiation) or unexpected parameters of the electron population in the jet, very different from \sft. Our model does not require any assumptions on these quantities, as the input on the X-rays, which serve as target photons, is from observation. 

Other explanations of the lack of jet signatures could be in the macroscopic picture, such as a possible intermittent nature of the jet, effects of an off-axis line of sight, a larger efficiency of energy dissipation in the collision region, or an unusual jet geometry -- as there may be effects from jet recollimation, twisting, or precession. Propagation effects could explain the suppression of PeV gamma rays, which may be re-processed in the source or absorbed in the extragalactic background light. More information on a possible jet might be obtained by very long baseline radio interferometry, or by late-term radio observations \cite{Generozov:2016oon} if the observed radio signal is interpreted as the afterglow of a relativistic jet; see also \cite{Alexander:2020xwb} for further discussions.

We have described the observation of a neutrino coincident with the tidal disruption event AT2019dsg in a jetted TDE model. In our interpretation, the unusually high X-ray luminosity  of \dsg\ is the reason for the efficient \n\ production, which implies that  X-ray-bright \tds\ might also be neutrino-bright. 
We have also shown that the late time of the neutrino signal (about 150 days after the optical peak) is not a coincidence if the neutrino production radius scales with the black body radius, whereas a very early neutrino signal close to the peak is not expected because the X-rays have not isotropized yet.  
Energetics and parameters match a unified TDE model from numerical simulations \cite{Dai:2018jbr}, which have led to our concordance model.  
 Indeed, for a large enough black hole spin, a jet may be expected in a unified picture of \tds\ in addition to a mildly relativistic outflow~\cite{Dai:2018jbr}, which has been (indirectly) observed~\cite{Stein:2020tem}.  A preliminary support to the jet hypothesis also comes from  optical polarimetry of \dsg\ \cite{Lee_2020}.
 
If \dsg\ and \sft\ are both jetted \tds, then one will have to conclude that the phenomenology of \tds\ is very diverse.
New dedicated studies will be needed to
explain this variety in terms of parameters such as  the black hole mass and spin, the type of disrupted star, the type of star-black hole approach trajectory, and the spectral energy distribution. The diversity will then impact the estimate of the diffuse flux of \ns\ from \tds. Our calculations show that  \dsg-like \tds\ could contribute  to the total \n\ flux observed at \ic\ at the per-cent level.  

We conclude that TDEs might be a promising class of neutrino emitters. While we have presented only one model here, other possibilities are conceivable, such as the interaction of an isotropic outflow with UV photons~\cite{Stein:2020tem},  non-relativistic shocks forming in the environment \cite{Fang:2020bkm},   or a neutrino production from the accretion disk itself, especially radiatively 
inefficient accretion flows or  magnetically arrested disk states~\cite{Hayasaki:2019kjy}; the neutrino production may also happen in a hot corona similar to that of an Active Galactic Nucleus (AGN)~\cite{Murase:2020lnu}. Our model is unique in that we have emphasized the connection to the X-ray observations, we have described the late, post-peak neutrino observation, and we have obtained a sufficiently high neutrino fluence to describe the observations in spite of a relatively small assumed SMBH mass.


{\bf Acknowledgments.} We would like to thank Anna Franckowiak, Marek Kowalski, Robert Stein, Andrew Taylor and Sjoert van Velzen for useful discussions. This work has been supported by the  European Research Council (ERC) under the European Unions Horizon 2020 research and innovation programme (Grant No. 646623), and by the US National Science Foundation grant number PHY-1613708. 

{\bf Author contributions.} The theoretical ideas were equally developed by CL and WW, numerical simulations were performed by WW. The artwork was produced by CL, the results figures by WW. Both authors contributed equally to the manuscript writing.

{\bf Corresponding author.} 

Correspondence to Walter Winter.

{\bf Author information.}

$^1$Deutsches Elektronen-Synchrotron (DESY), Platanenallee 6, D-15738 Zeuthen, Germany

$^2$Department of Physics, Arizona State University,  450 E. Tyler Mall, Tempe, AZ 85287-1504, USA


\section*{Methods} 
\label{sec:methods}

A star (of mass $m$) is disrupted by a supermassive black hole (\bh, mass $M$) if (i) it falls within a distance  less than the  tidal radius:  
    \begin{equation}
r_t  \simeq  9 \,  10^{12} \, \mathrm{cm}~\left(\frac{M}{10^6 \msun} \right)^{1/3}\frac{R}{\rsun}\left( \frac{m}{\msun} \right)^{-1/3}  \nonumber
\label{rt}
\end{equation}
and (ii) the tidal radius exceeds the \bh\ \scw\ radius, 
\begin{equation}
    R_s  =  \frac{2 M G}{c^2}  \simeq  3 \, 10^{11} \, {\rm cm} \left( \frac{M}{10^6 \msun} \right) \, ,
\end{equation}
as otherwise the star is swallowed by the black hole as a whole. The latter condition 
implies that the \bh\ mass is bounded from above by the Hill's mass \cite{Hills75}: $M \lesssim M_H \simeq 4 \, 10^{7}~\msun$, for a solar-type star being disrupted (see also \cite{Kochanek:2016zzg} for a more detailed discussion based on \td\ demographics). 
When modeling a \td\ emission, an upper bound on the total energy is given by the rest energy of the disrupted star, $E_{\mathrm{max}} \sim \msun c^2 \simeq 1.8 \times 10^{54}~{\rm erg}$ for a solar-mass star. 
A useful benchmark parameter is the \bh\  Eddington luminosity: 
 $L_{\mathrm{Edd}}\simeq 1.3~10^{44}~{\rm erg/s} \left(M/(10^6 \msun) \right)$.
 
The Blandford-Znajek mechanism \cite{1977MNRAS.179..433B} suggests that  a weak initial magnetic field in the accretion disk in combination with a high black hole spin can lead to the formation of a jet. 
 Numerical simulations of \tds\ that are based on general relativistic radiation magnetohydrodynamics confirm this hypothesis; see, in particular, the unified model in  \cite{Dai:2018jbr}, where a relatively high spin and $M=5\, 10^6 \msun$ were used. 
 This simulation obtains an average mass accretion rate (at near-peak times) $\dot M \sim 10^2 L_{\mathrm{Edd}}$ (see also \cite{DeColle:2012np,Guillochon:2012uc}), of which $ \sim 20\%$ and $\sim 3\%$ go into the jet and the bolometric luminosity, respectively (a remaining 20\% powers the outflow). These fractions result in a moderately super-Eddington jet, 
 and a total radiative emission near the Eddington limit (assuming  the results of \cite{Dai:2018jbr} can be rescaled  for black holes of different masses):
 \begin{eqnarray}
L^{\mathrm{phys}}_{\mathrm{jet}} \simeq 20 \, L_{\mathrm{Edd}} \simeq  3~10^{45}~\mathrm{\frac{erg}{s}} \left(\frac{M}{10^6 \msun}\right); \label{equ:daiscalingsjet} \\ 
L_{\mathrm{bol}} \simeq 3 L_{\mathrm{Edd}} \simeq 4~10^{44}~\mathrm{ \frac{erg}{s}} \left(\frac{M}{10^6 \msun}\right)~. 
\label{equ:daiscalingsbol}
\end{eqnarray}

In \cite{Dai:2018jbr}, the density profile of the accretion disk was modeled, 
indicating that the typical  size of the optically thick region 
(i.e., the radial distance where the optical depth for electron scattering is equal to unity)  
is, 
 \begin{eqnarray}
R_{\mathrm{BB}} \simeq 10^3 \, R_S  \simeq 3\, 10^{14} \,  \mathrm{cm} \left(\frac{M}{10^6 \msun}\right) ~,
\label{equ:daiscalings2}
\end{eqnarray}
approximately, and for $M \sim 10^6 \msun$ (the validity over wide ranges of $M$ has not been studied).

The velocity profile of the gas indicated increasingly fast outflows in regions of decreasing density (away from the plane of the accretion disk and closer to the jet), with speeds reaching $v\simeq 0.1~ {\mathrm{c}}$ or even $v\simeq 0.5~ {\mathrm{c}}$.   

From a comparison with the measured parameters of \dsg, an overall consistency appears. We note in particular the good agreement of the blackbody luminosity and radius 
(or, in other words, of the measured blackbody luminosity and temperature, via the Stefan-Boltzmann law, from which we find
$R_{BB}= \left(L_{BB}/4\pi \sigma_{SB} T^4_{BB} \right)^{1/2}\sim 4~10^{14}$ cm, in agreement with the value quoted in \cite{vanVelzen:2020cwu})
with Eqs. (\ref{equ:daiscalingsbol}) and (\ref{equ:daiscalings2}) 
, which indicate a black hole mass $M\simeq 10^6\msun$ for \dsg.
This value of $M$ also ensures basic consistency with the measured X-ray emission radius, which is found to be up to a factor of a few larger than $R_s$ (see, e.g., the NICER measurement, $r_X=6.8(+0.9,-0.7)~ 10^{11}$ cm, {[}Ref. \cite{2019ATel12825....1P}{]}, and might be underestimated due to observational effects  \cite{Stein:2020tem}.   
We note that higher values of the \bh\ mass, $M\sim 10^7 \, M_\odot$, are obtained using the empirical SMBH -- galaxy bulge mass correlation; see, for example, \cite{McConnell:2012hz}; however, a TDE-specific relationship, which includes TDE demographics, indicates scattering around $M\simeq 10^6\msun$ for that bulge mass \cite{Wevers:2019tzk}. A new method based on TDE dynamics \cite{Ryu:2020gxf}, gives at estimate of  $M \simeq 1.3 \,  10^6 \, M_\odot$ for \dsg, consistently with our choice. We 
stress that, should a higher black hole mass be established in the future, our model would remain valid, although with modified parameters. For instance, one may assume that a smaller fraction of the total energy goes into the jet (at the expense of a physical jet luminosity below the Eddington luminosity), or a corresponding increase of the physical jet luminosity potentially coming with enhanced neutrino production (at the expense of increased tension with signatures of the jet).
    
Moving now to describing the long term evolution ($t-t_{\rm peak}\gtrsim 10$ days) of a \td\ signal, we note that 
no detailed numerical modeling exists, so far.
Therefore, this part of the signal is more open to speculation and variety of interpretation. 
Here we adopt $L_{\mathrm{BB}}$
as a quantity of particular relevance, as it is probably a direct indicator of  the accretion disk formed by the debris of the disrupted star.
We model the time evolution of $L_{\mathrm{BB}}$ following \cite{vanVelzen:2020cwu}, with a change from faster to slower cooling at  $t-t_{\mathrm{peak}}\gtrsim 100$ days (see Fig.~\ref{fig:lumi_jetted}),  
where the cooling rate approaches the trend  $L_{\mathrm{BB}} \propto t^{-5/3}$, which is expected if the mass accretion rate follows the fallback time of stellar material\cite{Hills75,Rees:1988bf},
 see dotted curve in Fig.~\ref{fig:lumi_jetted}.

Our proposed jetted \td\ scenario builds on the methods of  \cite{Lunardini:2016xwi}. In the remainder of this section, the main features,  assumptions and inputs are described:

{\bf (i) Jet variability, Lorentz factor and physical energy. } 
For the jet, a bulk Lorentz factor $\Gamma \sim {\mathcal O}(10)$ is a natural value, inspired by AGN observations (e.g. \cite{Chai:2012ns}) and consistent with the best known jetted \td, \sft\ \cite{Burrows:2011dn}. We take $\Gamma = 7$,  and assume a viewing angle zero, therefore the Doppler factor is $D=2 \, \Gamma \simeq 14$; these values are centered around the usual assumption of a boost factor of about 10. Consequently, the jet opening angle can be estimated as $1/\Gamma$. Matter propagating in the jet has density and velocity inhomogeneities, leading to collisions of plasma shells at the collision radius $R_C$ where internal shocks form, and proton acceleration and subsequent  \n\ production via proton-photon scattering occurs. The inhomogeneities are characterized by the variability time scale of the jet, $t_v$, for which the \scw\ time is a plausible lower limit: $t_v \gtrsim \tau_s \sim 2 \pi R_s/c \simeq 63 \, \mathrm{s} \, \left( M/(10^6 \msun) \right)$. A comparable value, $t_v \simeq 100$ s, was favored by the \sft\ data \cite{Burrows:2011dn}, and is adopted here.  Using the estimates above for $\Gamma$ and $t_v$, one obtains a typical $R_C \sim 2 \, \Gamma^2 \, c \, t_v \gtrsim 2 \Gamma^2 \, c \, \tau_S \sim \mathrm{few}\, \times 10^{14} \, \mathrm{cm}$. Note how this value is comparable to $R_{\mathrm{ BB}}$, \equ{daiscalings2}. 

For the physical energy of the jet, we assume $L^{\mathrm{phys}}_{\mathrm{jet}} = 3 \, 10^{45} \,  \mathrm{erg \, s^{-1}} \simeq 20 \, L_{\mathrm{Edd}}$ at peak time, in consistency with \equ{daiscalingsjet}.  We also assume that $L^{\mathrm{phys}}_{\mathrm{jet}}$ evolves with time proportionally to $L_{\mathrm{BB}}$ until when  $L^{\mathrm{phys}}_{\mathrm{jet}}$  drops below $L_{\mathrm{Edd}}$ and the jet is expected to cease \cite{Rees:1988bf}. The time of jet cessation depends on the (uncertain) evolution of $R_{\mathrm{BB}}$, and can take place at $\sim$170 to 300 days post-peak (see Fig.~\ref{fig:lumi_jetted}) -- which is in any case after the time of the \n\ detection; we apply an exponential cutoff $\propto \exp(-L_{\mathrm{edd}}/L_{\mathrm{jet}}^{\mathrm{phys}})$ to  the proton luminosity to include this effect. 
It can be estimated that over this time-scale, a total emitted energy $E_{\mathrm{tot}} \lesssim 3~ 10^{-2}~ \msun  c^2 $ is needed to power the jet.   
We assume that electromagnetic signatures of the jet cannot be seen due to absorption, similarly to the case of X-rays (discussed below).

{\bf (ii) Collision radius.} The long delay of the \n\ detection with respect to $t_{\mathrm{peak}}$ suggests little or no decrease of the \n\ luminosity over more than 100~days. To reproduce this feature,  we introduce a new element,  a time-decreasing collision radius $R_{\mathrm{C}}$.  
In particular, inspired by the numerical similarity $R_C \sim R_{\mathrm{BB}}$ at peak time,  we assume that $R_C$  follows the observed evolution of $R_{\mathrm{BB}}$ \cite{vanVelzen:2020cwu}:
\begin{equation}
 \frac{R_C}{10^{14} \, \mathrm{cm}} \simeq \left\{ 
 \begin{array}{ll}
 5.0 \, \exp \left( - \frac{t-t_{\mathrm{peak}}}{109 \, \mathrm{d}} \right) & , \,   t-t_{\mathrm{peak}} \le 150 \, \mathrm{d}
 \\
 1.3  & , \,  t-t_{\mathrm{peak}} > 150 \, \mathrm{d} \, .
 \end{array}
 \right.
 \label{equ:rscale}
\end{equation}
Generally, a time-decreasing $R_C$ can be justified in the context of the overall decline of the power of the jet, which might result e.g., in a decreasing value of $\Gamma$. 
We note that the estimate $R_C \sim 2 \, \Gamma^2 \, t_v $ does not literally hold in multi-zone collision models, but rather a more physical description of the collision radius should be done in terms of the distance between the plasma shells and their width, see  \cite{Bustamante:2016wpu} for a in-depth discussion. 
Since the pion production efficiency scales $\propto R_C^{-2}$, the drop in $R_C$ will enhance the late-term neutrino production.

{\bf (iii) Target photons.} Another key element of our model is that the background photons necessary for the photo-pion production originate \emph{externally} to the jet, as the X-rays that are emitted from the inner accretion disk (at $R \sim R_X$) are then back-scattered into the jet funnel, see Fig.~\ref{fig:cartoon}, right panel. This assumption is attractive because it links the \n\ production to \dsg\ being particularly bright in X-rays.
The description also naturally fits the neutrino energy, as the target photon energy to produce PeV neutrinos can be estimated (for external photons boosted into the jet frame)  as 
\begin{equation}
E_X \, [\mathrm{keV}] \simeq \frac{0.025}{ E_\nu \, [\mathrm{PeV}]} \, . \label{equ:nuenergy}
\end{equation}
Therefore, for the jetted TDE scenario with external radiation,  X-rays with the observed temperature are the ideal target.
In principle, some UV photons could also reach the collision region and serve as targets for neutrino productions. However, their contribution should be negligible, because: (i) 
the flux of back-scattered UV photons should  be small, since observations are consistent with an unabsorbed UV flux and the emission geometry is different -- see (vi) for their potential impact; and  (ii) unscattered UV photons would enter the (relativistically moving) collision region from behind, resulting in decreasing photo-pion efficiency.

The observed exponential decline of $L_X$  suggests that a time-dependent absorption effect might be at play. 
Hence, we consider a scenario where an expanding outflow obscures the X-rays.
 For an expansion speed  $v\simeq 0.1 \, c$ -- which is conservative, values reaching $v\simeq 0.5 \, c$  are expected closer to the jet, see \cite{Dai:2018jbr} -- 
we find that, over the characteristic X-ray decline time of $\sim 10$ days,  the cocoon expands out to at least a distance $\simeq  3 \, 10^{15} \, \mathrm{cm}$, 
which can serve as an estimate for the absorption radius  $R_{\mathrm{abs}}$.
Considering that $R_{\mathrm{abs}}$ exceeds the initial value of $R_{\mathrm{C}}$ by nearly an order of magnitude, 
it is realistic to expect that a fraction of the X-ray photons will be  absorbed/reprocessed over the length scale $R_{\mathrm{C}} \sim R_{\mathrm{BB}}$. The scattered photons will then serve as an external target photon field of isotropized X-rays, leading to Doppler-boosted (by a factor $D^2$, leading to enhanced pion production) target photon density similar to external photon targets in AGN, see \eg\ \cite{Murase:2014foa}, whose description we follow here. 

One can check that our proposed mechanism for the photon background is compatible with theoretical outflow models. Considering that in such models the matter density has a somewhat complicated dependence on radial distance and on time, and that several processes contribute to the scattering and absorption of X-rays~\cite{Roth:2015nza}, only a rough estimate can be presented here. The Thomson optical depth (from charge neutrality, assuming that the electron density is half of that of protons/neutrons and there is a significant contribution from free electrons) is given by
\begin{equation}
    \tau_T \simeq 2 \left( \frac{\rho}{10^{-14} \, \mathrm{g \, cm^{-3}}} \right) \left(\frac{ d }{10^{15} \, \mathrm{cm}} \, \right)~,
\end{equation}
where $d$ is the travelled distance.
In \cite{Dai:2018jbr} (see Fig. 4 there, for the densities, see Fig. 3), the numerically-calculated region where
the electron-scattering optical depth approaches 1 is shown, and its size is found to be  comparable with $R_{\mathrm{BB}}$ (for a wide range of angular distances from the jet funnel),
which justifies our assumption of photon isotropization at and beyond that scale of length. Note that the photon absorption opacity especially increases beyond the photo-ionization threshold of hydrogen (13.6 eV), which means that it is plausible that UV photons can escape whereas X-rays are confined; details are model-dependent.

To implement the scenario described above quantitatively,  
we model the unattenuated X-rays luminosity  according to  simulations for \tds\ with  slim disks, e.g.~\cite{Wen:2020cpm}, 
which show that the X-ray luminosity does not follow the mass fallback rate, but stays nearly constant up to the time of flare cessation (the mass accretion rate becomes sub-Eddington).  In \cite{Wen:2020cpm} an exponential drop over a time scale of 200 days post peak  is found for the \bh\ mass used here, which we incorporate into our model, assuming that the unattenuated light curve is at the level of the observations at the earliest times measurements are available ($t-t_{\mathrm{peak}}=17 \, \mathrm{d}$).
Note that the applicability of the slim disk model may be limited, especially at early times \cite{Wen:2020cpm}.
However, our neutrino light curve does not qualitatively change even if the X-ray luminosity follows the blackbody one.
We furthermore assume  that 10\% of the unattenuated X-rays isotropize and build up on the attenuation timescale, with the same energy spectrum as the unattenuated parent photon flux (which is plausible considering the relatively low rate or photon re-processing).  Note that this radiation will not be observable,  so  any  late-term  X-rays bounds  only  apply  to  the  thick blue (dashed) curve in Fig. \ref{fig:lumi_jetted}.

{\bf (iv) Hadronic content of the jet.}
 Protons are assumed to be accelerated at the collision radius $R_C$ by internal shocks to a power law spectrum $\propto E'^{-2}_p$ (primed indices refer to the shock frame) with a maximal energy determined by balancing the acceleration rate $t'^{-1}_{\mathrm{acc}}=\eta c / R'_L$ (with moderate $\eta =0.01$ and $R'_L$ the Larmor radius of the proton) with the synchrotron loss and an dynamical rates (so the Hillas criterium is satisfied). As the interactions occur in the optically thin (to $p\gamma$ interactions) regime, the requirements for proton acceleration are moderate. The (isotropic-equivalent) proton luminosity  is given by $L_p^{\mathrm{iso}} \simeq (2 \, \Gamma^2) \, \varepsilon\, L^{\mathrm{phys}}_{\mathrm{jet}}$,  see Fig. \ref{fig:lumi_jetted}, where $(2 \, \Gamma^2)$ is the beaming factor and  $\varepsilon$  is the transfer (dissipation) efficiency from jet kinetic energy into non-thermal radiation dominated by baryons. We take $\varepsilon \simeq 0.2$, which is 
  well within the range of typical values for Gamma-Ray Burst internal shock scenarios, see \eg\ \cite{Peer:2015eek,Sari:1997kn,Kino:2004uf,Bustamante:2016wpu,Rudolph:2019ccl}. 
 
Note that, if the non-thermal radiation from the jet  can escape, its non-observation results in a lower limit on the jet baryonic loading, $\xi_b$, which is usually defined as the ration of proton and electron injection luminosities.
 More specifically, assuming that the electrons are in equilibrium with the X-rays, one can estimate the minimal baryonic loading by comparing the required isotropic-equivalent proton luminosity to the X-ray bounds. Using  the late-term bound $L_X \lesssim 10^{42} \, \mathrm{erg \, s^{-1}}$~{[}Ref. \cite{vanVelzen:2020cwu}{]} one obtains $\xi_b \gtrsim 10^3 - 10^4$, which implies that the expected non-thermal X-ray signal is below the current bound if there is sufficiently enough energy in protons compared to electrons.  Comparable or larger numbers are typical for hadronic models of AGN, such as the first identified \n\ emitter, TXS 0506+056 \cite{Gao:2018mnu}.
 One should consider, however, that constraints depend on the photon spectrum of AT2019dsg. For example, if a AGN-like spectrum is assumed --  with its characteristic suppression in the X-ray band and likely peak the optical/UV and gamma-ray ranges -- the constraint on $\xi_b$ may be about an order of magnitude weaker.  One may wonder about the contribution of non-thermal X-rays from the jet to the \n\ production. Because these photons are produced in the jet co-moving frame (and so are not Lorentz-boosted), for them the \n\ production efficiency is lower compared to the external (boosted) photons; this efficiency problem has been studied for  TXS 0506+056 and AGN in general~\cite{Keivani:2018rnh,Rodrigues:2018tku}.
 
{\bf (v) Magnetic field and other assumptions. }
We assume that the magnetic field energy density takes a fraction of 10\% of the proton energy density~\cite{Lunardini:2016xwi} (corresponding to 2\% of the jet kinetic energy), which leads to a magnetic field $B' \simeq 90 \, \mathrm{G}$. 
The neutrino mixing angles are taken from from NuFIT 4.1 (2019)~\cite{Esteban:2018azc}. Note that in our model, the information from radio data are not directly relevant for the \n\ production, mainly because $R_{\mathrm{radio}}\gg R_C$. 

 Our numerical treatment closely follows \cite{Lunardini:2016xwi}, with the time evolution of the spectra being calculated in discrete steps of 1 day width.

 {\bf (vi) Shape of the resulting neutrino spectrum and light curve. } 
The neutrino light curve (Fig.~\ref{fig:lumi_jetted}) is approximately described by $L_\nu \propto L_p^{\mathrm{iso}} \, f_\nu$, where $f_\nu \propto L_X^{\text{isotropized}}/R_C^2$ is the neutrino production efficiency and $ L_X^{\text{isotropized}}$ is the luminosity of the isotropized X-rays. Consequently, there are three competing processes determining its shape: (i) the evolution of $L_p^{\mathrm{iso}}$, dropping with the BB luminosity, see Fig.~\ref{fig:lumi_jetted}, (ii) the decrease of $R_C$ following \equ{rscale},
and, (iii) the evolution of $ L_X^{\text{isotropized}}$.
 The time of the first peak of the neutrino lightcurve is determined by the timescale of isotropization of the X-rays, which is inferred from the observed obscuration timescale. 
The decline following the first peak comes from the combined effect of decreasing proton luminosity -- including the break at $t-t_{\mathrm{peak}} \sim  100 \, \mathrm{d}$ --
and of the decreasing collision radius. The peak at $t-t_{\text{peak}}\sim 150 \, \mathrm{d}$ originates from the stagnation of the collision radius decrease in \equ{rscale}. The later decline, after the second peak, follows the decrease $L_p^{\mathrm{iso}}$ and its eventual  exponential cutoff, occurring when $L^{\mathrm{phys}}_{\mathrm{jet}}$ falls below the Eddington luminosity (at $t-t_{\mathrm{peak}} \sim  300 \, \mathrm{d}$, shown in Fig.~\ref{fig:lumi_jetted}). Note that, following the slim disk model, the unattenuated X-ray luminosity is assumed to decline exponentially over a timescale of 200 days. The effect of that model-dependent time evolution on the neutrino lightcurve is secondary:
As a test, we produced results for an alternative scenario, where the unattenuated X-ray luminosity evolves with time following the BB luminosity; we found that the qualitative conclusions are robust. 

Let us now comment on the neutrino spectrum in Fig.~\ref{fig:fl_jetted}. Its general form roughly matches the naive estimate presented in \equ{nuenergy} for the $\Delta$-resonance, however its detailed shape is broader, because multi-pion production is efficient due to the high target photon temperature.
A number of uncertainties affect the neutrino spectrum and lightcurve. Specifically, the high energy end of the spectrum depends on the maximal proton energy (determined by the acceleration efficiency), $E_{p,\mathrm{max}} \gtrsim 20~ E_\nu \simeq 4 \, \mathrm{PeV}$ (estimated from the observed neutrino energy $E_\nu \simeq 0.2 \, \mathrm{PeV}$). 
The neutrino spectrum  would be enhanced at high energy for higher $E_{p,\mathrm{max}}$, if a background of lower energy photon targets were available, for example as isotropized UV photons. In such a case, 
the neutrino spectrum would peak at about a factor of twenty higher energy according to \equ{nuenergy}, which implies that  $E_{p,\mathrm{max}} \gtrsim 80 \, \mathrm{PeV}$ is required. 
Conservatively, we did not include this possibility in the main text as there is no observational evidence for UV obscuration/back-scattering.  Moreover, due to the different sizes of the X-ray and UV photospheres, one can not apply the same assumptions as for the X-rays back-scattering (thus introducing further uncertainty).  Still, we have done that in a test run -- assuming the same acceleration efficiency as in the main text -- with the goal of studying an extreme scenario.  
As expected, we find an overall more energetic neutrino spectrum, which would be less compatible with the observed neutrino energy. 
Note that un-scattered UV photons or X-rays (which may be relevant at early times, before obscuration sets in) have a $\sim 10^2$ lower Doppler factor than  back-scattered photons, since they come from behind the production region. Therefore their contribution to the neutrino flux is suppressed.

{\bf Data availability.} The  data  that  support  the  plots  within  this  paper  and  other  findings  of  this  study  are  available in the supplementary information or from the corresponding author upon reasonable request.

{\bf Code availability.} The codes that support the plots within this paper and other findings of this study are available from the corresponding author upon reasonable request.

{\bf Competing interests.} The authors declare that there are no competing interests.


\clearpage





\end{document}